\documentstyle[12pt]{article}

\begin{document}

\author{Alexander V. Shchagin\thanks{%
E-mail: shchagin@kipt.kharkov.ua}}
\title{Diffraction on a polycrystal for investigations and diagnostics of X-ray
radiation of relativistic particles in a forward direction }
\date{Kharkov Institute of Physics and Technology, Kharkov 61108, Ukraine }
\maketitle

\begin{abstract}
It is shown that the diffraction on a polycrystal can be used for
investigation and diagnostics of X-ray radiation emitted in a forward
direction by relativistic charged particles moving in crystalline or other
targets or fields. Methods for measuring radiation spectral density,
divergence, and linear polarization at any requisite energy from a few units
to tens of keV are proposed. The explanation for the origination of
experimentally observed and earlier unidentified spectral peaks as a result
of Bragg diffraction on a polycrystal is described. The experiment for
verification of the explanation is suggested.

.


.

\end{abstract}

\section{Introduction}

There exist several kinds of X-ray radiation of relativistic charged
particles moving through a radiator, i.e., amorphous or crystalline targets
or fields. These are ordinary bremsstrahlung, transition radiation (TR),
resonance transition radiation, coherent bremsstrahlung, channeling
radiation, parametric X-ray radiation (PXR), undulator radiation, Thomson or
Compton scattering, etc. Most of them are going from the radiator in a
forward direction mainly within the angle of about $\gamma ^{-1}$ relative
to the incident particle velocity vector $\overrightarrow{V}$, where $\gamma 
$ is the relativistic factor of incident particles. Several of the
above-mentioned kinds of radiation can be generated in the radiator (e.g.,
in a crystal) simultaneously, and the total radiation in a forward direction
(RFD) is going along the vector $\overrightarrow{V}$. To investigate the
composition of the total radiation and the role of its components,
experimental measurements of the RFD spectral properties are needed.
However, it is generally difficult to measure the spectral properties of the
RFD from relativistic particle beams on account of its high intensity, wide
spectrum and a restricted counting rate of spectrometers. In the gamma-ray
band, it is the Compton scattering that is successfully used for
investigations of the RFD. In the X-ray band, the use of Bragg diffraction
seems to be more natural. In this paper we suggest that the Bragg
diffraction in a polycrystal placed behind the radiator should be used for
measurements of the RFD spectral density, divergence and linear polarization
at any wanted X-ray energy. Besides, we shall discuss the experiment, where,
in our opinion, the Bragg diffraction of X-ray RFD by the polycrystal was
observed.

\section{How to measure X-ray radiation in a forward direction with the use
of a polycrystal}

The scheme of the setup for measurements of the X-ray RFD properties is
shown in Fig. 1. The beam of incident relativistic particles from the
accelerator passes through the radiator R and generates RFD going along the
particle beam within the angle of about $\gamma ^{-1}$. Then the RFD crosses
the polycrystalline foil P. The particle beam can be either deflected by the
bending magnet or can pass through the foil, too, if its radiation in the
foil does not prevent the observation of radiation under study. The
spectrometric X-ray detector D is installed at an observation angle $\theta $%
. The polycrystalline foil consists of a number of randomly aligned
crystalline grains. Some of them can appear oriented relative to the vector $%
\overrightarrow{V}$ and the observation direction $\overrightarrow{\Omega }$%
to satisfy the Bragg condition. One of these grains with the
crystallographic planes denoted by the reciprocal lattice vector $%
\overrightarrow{g}$ is shown in the foil in Fig. 1. The X-ray RFD of Bragg
energy $E_B$ will be reflected by these planes into the detector. Thus, the
spectrometric detector will be able to register several spectral peaks of
energies $E_B,$ corresponding to several main crystallographic planes of the
polycrystal. These energies can be found by the formula from Ref. \cite{sh}

\begin{equation}
\label{ebr}E_B=\frac{c\hbar g^2}{2\left| \overrightarrow{g}\overrightarrow{%
\Omega }\right| }, 
\end{equation}
where $\overrightarrow{\Omega }$ is the unit vector in the observation
direction at the observation angle $\theta $ relative to the particle
velocity vector $\overrightarrow{V}$ and the RFD axis, $\left| 
\overrightarrow{g}\overrightarrow{\Omega }\right| =\left| \overrightarrow{g}%
\right| \cos \frac{\pi -\theta }2=g\sin \frac \theta 2$, $g=\left| 
\overrightarrow{g}\right| =\frac{2\pi \sqrt{l^2+m^2+n^2}}a$, $a$ is the
lattice constant; $l,m,n$ are the Miller indices for the crystallographic
planes with a nonzero structure factor.

The scheme in Fig. 1 is similar to the one used in the well-known
Debye-Scherrer method for investigations of polycrystalline samples by a
monochromatic X-ray beam. Here, we propose that this scheme with the known
polycrystal should be used for investigation and diagnostics of intense RFD
X-ray beams that may have wide and complicated spectra.

\subsection{Spectral density}

The number of counts registered by the detector in the spectral peak of
energy $E_B$ is proportional to the RFD spectral density at this energy. For
absolute measurements of RFD spectral density, one should calibrate the
foil+detector system using the radiator with the known spectral density of
X-ray radiation and provide a reliable monitoring of the number of incident
electrons. For measurement of RFD spectral density in arbitrary units, one
can change radiators or their properties at a fixed geometry of both the
foil and the detector and register number of counts in the spectral peaks at
energy $E_B$ generated by constant number of incident particles. For
example, in this way one can measure the spectral density of channeling
radiation as a function of crystal-radiator alignment. Note that only a
small part of RFD can be diffracted by a thin polycrystalline foil into the
detector. This is favorable for preventing the spectrometer from overloading
at measurements of intense X-ray RFD beams. The wanted energy $E_B$ can be
provided by a proper choice of the polycrystal and the observation angle in
accordance with the formula (\ref{ebr}).

\subsection{Divergence}

The width $\Delta E$ of the measured spectral peak at energy $E_B$ is a
function of the experimental angular resolution $\Delta \theta $ and the
incident X-ray RFD beam divergence $\alpha $ in the observation plane, and
also the energy resolution of the detector $\Delta E_d$. The experimental
angular resolution is determined by the angular size of both the detector
and the RFD beam spot on the foil in the observation plane. Using the Eq. (%
\ref{ebr}), one can find

\begin{equation}
\label{de}\Delta E^2=\Delta E_d^2+\left( \frac{E_B}{2\tan \frac \theta 2}%
\right) ^2\Delta \theta ^2+\left( \frac{E_B}{2\tan \frac \theta 2}\right)
^2\alpha ^2. 
\end{equation}
In practice, the divergence can be measured provided that the spectral peak
broadening due to the RFD divergence exceeds or is comparable to the
broadening due to the experimental angular resolution and energy resolution
of the detector. In this case, the divergence $\alpha $ in the observation
plane can be found from (\ref{de}).

\subsection{Linear polarization}

The Bragg diffraction intensity has its maximum for X-rays polarized in the
plane perpendicular to the diffraction plane, and its minimum for the X-rays
polarized in the plane of diffraction. These maximum and minimum are
particularly pronounced at $\theta $ close to $\pi /2$. Due to these
well-known peculiarities of Bragg diffraction, the setup shown in Fig. 1
should possess the polarization analyzing power. The setup can be used for
measurements of the RFD linear polarization at energy $E_B$. To this end,
one should perform measurements of polycrystal-diffracted radiation at a
fixed observation angle $\theta $ as a function of the azimuthal angle of
the detector rotation around the vector $\overrightarrow{V}$.

\section{Discussion of some experimental results from Ref. [2]}

In our opinion, the diffraction of RFD by the polycrystal has been observed
in Ref. \cite{jap}. The experimental setup in Ref. \cite{jap} was partially
similar to the one shown in Fig. 1. The authors of Ref. \cite{jap} studied
the PXR and the diffracted TR in the Bragg direction, the radiations being
generated by the 150 MeV electron beam in silicon single-crystal radiators
of various configurations. Diffraction of TR realized in the same
single-crystal radiators. Behind the radiator, a 10 $\mu m$ thick molybdenum
foil was installed. The characteristic X-ray radiation, excited in the foil,
was used for monitoring the number of beam electrons that have passed
through the radiators. Those authors have measured a series of nice spectra
having a low spectral background by a Si(Li) detector at $\theta
=25.8^{\circ }$. In the spectra they observed clearly marked spectral peaks
of PXR and diffracted TR from the radiator, and also the peaks of
characteristic X-ray radiation from the molybdenum foil at reference
energies $E_{K\alpha }=17.45$ keV and $E_{K\beta }=19.6$ keV. Besides, they
observed spectral peaks with energies $\approx 12.5$ keV and $\approx 25.0$
keV, the origin of which was not identified in Ref. \cite{jap}. Here, we
shall discuss the data concerned with these unidentified spectral peaks
(USPs).

In Ref. \cite{jap} the authors have noted that the molybdenum foil was
amorphous. To understand the origin of the USPs, let us suppose that the
molybdenum foil is polycrystalline. This polycrystal can diffract the
radiation of Bragg energies from the RFD, generated in the radiator, into
the cone and, in particular, in the detector direction. The Bragg energies
calculated by formula (\ref{ebr}) for the crystallographic planes with
nonzero structure factors (110), (220), (200) of molybdenum lattice \cite
{kitt} at the reference lattice constant $a$=3.15\AA\ and $\theta
=25.8^{\circ }$ are $E_B^{(110)}=12.5$ keV, $E_B^{(220)}=25.0$ keV, $%
E_B^{(200)}=17.6$ keV, respectively. The calculated energies $E_B^{(110)}$
and $E_B^{(220)}$ are practically coincident with the ones of both USPs
observed in Ref. \cite{jap}. The energy $E_B^{(200)}$ is close to $%
E_{K\alpha }$ of the characteristic peak, and these peaks are not seen
resolved in the experimental spectra in Ref. \cite{jap}.

Consider some other experimental data concerned with USPs described in Ref. 
\cite{jap}:

i. The USPs disappear if the radiator is removed. This is because the RFD
from the radiator disappears. Therefore, only characteristic peaks excited
by the electron beam in the foil are seen in Fig. 11b of Ref. \cite{jap}.
Note that the spectral peaks of PXR with the energies practically equal to
the energies of USPs can be generated by the beam electrons in molybdenum
grains. Their absence in Fig. 11b of Ref. \cite{jap} means that the PXR from
the polycrystal is weak and is no obstacle for correct measurements of RFD
(see item 3 in nest section).

ii. The USPs disappear if the molybdenum foil is removed. This is because
the Bragg diffraction without the polycrystal is absent.

iii. The energies of USPs do not vary with significant variations of the
crystal-radiator alignments. This is because the RFD is going along the
fixed vector $\overrightarrow{V}$ independently of the crystal-radiator
alignment.

iv. The energies of USPs are the same at arbitrary alignment of the
molybdenum foil. This is because the RFD is diffracted by the molybdenum
grains which appears at appropriate for Bragg diffraction alignment
independently of the alignment of the whole foil.

v. The 12.5 keV USP seems to vanish with the alignment of the Si
crystal-radiator $\left\langle 100\right\rangle $ axis close to the incident
particle beam axis (see Fig. 12c in Ref. \cite{jap}). This may be due to a
significant broadening of the USP as a result of an appreciably increased
RFD divergence. The increased RFD divergence may be a result of increased
electron beam scattering in the crystal-radiator at this alignment because
of the crystal-radiator configuration \cite{jap}. Besides, the increasing of
the electron beam scattering is possible at motion of electrons along the $%
\left\langle 100\right\rangle $ strings of the crystal.

Thus, the above-considered experimental data from Ref. \cite{jap} seems are
in agreement with our explanation of the USPs origin as a result of the RFD
Bragg diffraction by the molybdenum polycrystal.

\section{Results and discussion}

1. In this paper we have suggested the methods for diagnostics and
measurements of intense X-ray RFD. They permit measurements of spectral
density, divergence and linear polarization of the RFD with the use of Bragg
diffraction on a polycrystal. The methods seem relatively simple and
inexpensive, as only a single polycrystalline foil with an arbitrary
alignment should be installed, and ordinary spectrometric detector(s) can be
used for measurements at any energy chosen in the range from several keV to
tens of keV.

2. Here, we have suggested the explanation of spectral peak origination at
energies of about 12.5 and 25.0 keV, observed and unidentified in Ref. \cite
{jap}. The peaks are due to the Bragg diffraction of RFD from the radiator
by a polycrystalline molybdenum foil installed behind the radiator. This
explanation can be additionally verified with the experimental setup
described in Ref. \cite{jap}. For this purpose, one can vary the
registration angle of the detector $\theta $ and observe variations of the
spectral peak energies. For the molybdenum polycrystal, they should obey the
following formulae obtained from (\ref{ebr}):

\begin{equation}
\label{e11}E_B^{\left( 110\right) }=\frac{1.969\cdot \sqrt{2}}{\sin \frac
\theta 2}keV 
\end{equation}
for the spectral peak from the $\left( 110\right) $ plane of molybdenum ($%
\approx $12.5 keV in Ref. \cite{jap} at $\theta =25.8^{\circ }$) and

\begin{equation}
\label{e22}E_B^{\left( 220\right) }=\frac{1.969\cdot \sqrt{8}}{\sin \frac
\theta 2}keV 
\end{equation}
for the spectral peak from the $\left( 220\right) $ plane of molybdenum ($%
\approx $25.0 keV in Ref. \cite{jap} at $\theta =25.8^{\circ }$). Besides, a
new peak at energy

\begin{equation}
\label{e2}E_B^{\left( 200\right) }=\frac{1.969\cdot 2}{\sin \frac \theta 2}%
keV 
\end{equation}
from the $\left( 200\right) $ plane of molybdenum may appear. In the
experimental conditions \cite{jap} at $\theta =25.8^{\circ }$, the energy of
this spectral peak $E_B^{\left( 200\right) }=17.64$ keV is close to the one
for the characteristic $K\alpha $ peak at $E_{K\alpha }=$ $17.45$ keV. These
peaks could not be resolved by the detector with an energy resolution of 450
eV used in \cite{jap}.

Besides, our explanation can be verified by using another kind of
polycrystal at the same observation angle. For example, the copper
polycrystal can diffract the RFD with energies 13.3, 15.4, 21.8 keV from
crystallographic planes (111), (200), (220) respectively at $\theta
=25.8^{\circ }$.

3. As we mentioned above, the electron beam can generate the PXR in the
randomly aligned crystal grains of a polycrystal. However, only a small part
of these grains has a proper alignment and produces the PXR reflection in
the observation direction. One can estimate relative number of such grains.
As the angular size of PXR reflection is about $\gamma ^{-1}$, only grains
with reciprocal lattice vectors $\overrightarrow{g}$ within the solid angle $%
\sim \gamma ^{-2}$ can take part in generation of the reflection in fixed
observation direction. The relative number of such grains is $\sim \frac{%
\gamma ^{-2}}{2\pi }$ . The effective thickness $T_{eff}$ for generation of
PXR reflection in fixed observation direction of the polycrystal with
thickness $T$ may be estimated as

\begin{equation}
\label{tt}T_{eff}\approx \frac{\gamma ^{-2}}{2\pi }T. 
\end{equation}
Thus, the PXR from thin polycrystal should be weak and therefore is not seen
in Fig. 11b of Ref. \cite{jap}.

4. To investigate radiation in a forward direction in the wanted X-ray
energy range, one can use a polycrystal and position-sensitive X-ray
detector(s) installed at corresponding observation angles.

The diffraction on a polycrystal provides good possibility for studying the
PXR and/or other kinds of radiation diffracted in a crystal-radiator with
simultaneous measurements of radiation in a forward direction. For example,
the search for PXR in a forward direction may be continued and/or channeling
or transition radiation may be studied with the use of a polycrystal.

\section{Acknowledgments}

The author is thankful to V.M. Sanin for discussion of this paper. The paper
became possible partially due to Project STCU \# 1031 from Science and
Technology Center in Ukraine.

\section{Figure caption}

Fig. 1. The relativistic particle beam passes through the radiator R and
generates radiation in a forward direction (RFD). The RFD is going along the
particle velocity vector $\overrightarrow{V}$and passes through the
polycrystalline foil P. One of randomly aligned grains of the polycrystal
with the crystallographic planes and corresponding reciprocal vector $%
\overrightarrow{g}$ is shown in the foil. The spectrometric X-ray detector D
is installed at observation angle $\theta .$ The observation direction is
shown by the unit vector $\overrightarrow{\Omega }$. The detector can
register polycrystal-diffracted RFD at Bragg energies.

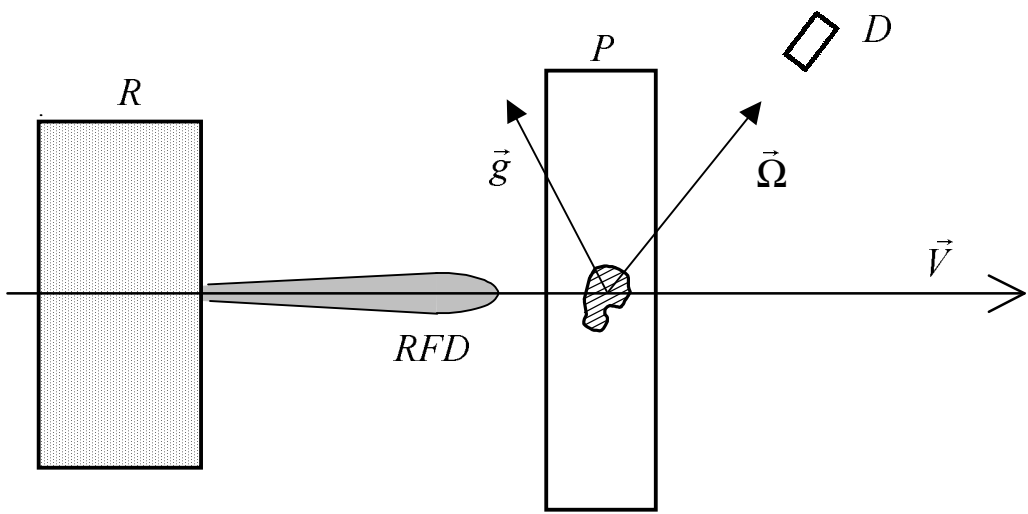
\end{document}